\documentclass[conference,usenatbib]{basi}
\usepackage[T1]{fontenc}
\usepackage[british]{babel}
\usepackage[varg]{txfonts}

\begin{document}
\title[Non-steady outflows and QPOs around black hole]{
Formation of non-steady outflows
and QPOs around black hole}
\author[S.~Das et~al.]%
       {Santabrata Das$^1$\thanks{email: \texttt{sbdas@iitg.ernet.in}},
       I.~Chattopadhyay$^{2}$ and Aunj Nandi$^{3}$\\
       $^1$Department of Physics, IIT Guwahati, Assam, India\\
       $^2$ARIES, Manora Peak, Nainital 263002, Uttarakhand, India\\
       $^3$Space Astronomy Group, SSIF/ISITE Campus, ISRO Satellite Centre, \\
           Outer Ring Road, Marathahalli, Bangalore, India}

\pubyear{2013}
\volume{8}
\pagerange{31--36}
\setcounter{page}{31}


\maketitle
\label{firstpage}

\begin{abstract}
We study the time dependent properties of sub-Keplerian viscous accretion
flow around the black holes. We find that rotating matter feels centrifugal
barrier on the way towards the black holes and eventually, shock transition
is triggered allowing a part of the post-shock matter to eject out as
bipolar outflow. This outflowing matters are supposed to be the precursor
of relativistic jets.  As viscosity is increased, shock becomes unsteady
and start to oscillate when viscosity reached its critical value. This
causes the inner part of the disk unsteady resulting periodic ejection 
of matter from the post-shock region. Also, the same hot and dense 
post-shock matter emits high energy radiation which modulates
quasi-periodically. The power density spectra confirms this features
as most of the power is concentrated at a narrow frequency range ---
a characteristics ($i. e.$ Quasi-Periodic Oscillation) 
commonly seen in several outbursting black hole candidates.
\end{abstract}


\section{Introduction}

Outflows/Jets are commonly seen in extreme gravitating systems ranging from
stellar mass to supper massive black holes. They are believed to be 
originated from the regions few tens of  
Schwarzschild radius (${\rm r_g}$) around the central objects. In addition,
\citet{gfp03} 
reported that quasi steady outflows are generally ejected
in the hard state which indicates that the generation of outflows 
are expected to depend on various states of the accretion disk.
Inner boundary condition for accretion onto black hole demands
that the accreting matter must be transonic and sub-Keplerian in
nature. Close to the black hole, accreting matters are slowed
down due to centrifugal pressure and eventually piled up to
form a torus like structure at the inner part of the disk.
Such a geometry behaves like an effective boundary layer to
the accreting matter from the outer edge where flow may
undergo steady or non-steady shock transition
\citep{c89,ttaf,cm95,c96,d07}. Usually,
the post-shock matter is hot and dense because of compression
and behaves as a source of high energy radiation \citep{ct95,cm06}. Also, due
to thermal pressure gradient, a part of the
accreting matter is deflected from the boundary layer and
ejected as bipolar outflow \citep{c99,dcnc01,cd07,dc08}.
When shock oscillates, the 
boundary layer becomes non-steady and effectively, both the
outflow and the emergent radiation flux modulate quasi periodically
in a similar manner. In this work, we study the shock
oscillation, outflow, and emitted flux variation in terms of the
flow parameters. In the next Section, we describe simulation method
and finnaly present results and discussion.

\section{Simulation}

We consider the time dependent axisymmetric 2-dimensional 
viscous accretion flow around a
Schwarzschild black hole and study the time evolution of flow
variables using smooth particle hydrodynamics (SPH) scheme
\citep{cm95}. We
approximate the space-time geometry around a non-rotating
black hole assuming the pseudo-Newtonian potential introduced
by \citet{pw80}. In this work, we adopt the dynamical
viscosity prescription from \citet{cm95} which
is given by,
$$
\eta = \nu \rho = \alpha \rho (a^2+v^2){\sqrt{2x(x-1)^2}} 
$$
where, $\nu$ is the kinematic viscosity, $\alpha$ is the dimensionless
viscosity parameter, $a~(=\sqrt{{\gamma P}/{\rho}})$ is the sound
speed, and $v=\sqrt{v^2_x+v^2_z}$ is the total velocity. Other variables
have their usual meanings.
In this work, the distance, velocity and time are measured in units of
$r_g=2GM_{BH}/c^2$, $c$ and $t_g=2GM_{BH}/c^3$, where $G$, $M_{BH}$ and $c$
are the gravitational constant, the mass of the black hole and the speed
of light, respectively.
%

\begin{figure}
\centerline{\includegraphics[width=8cm]{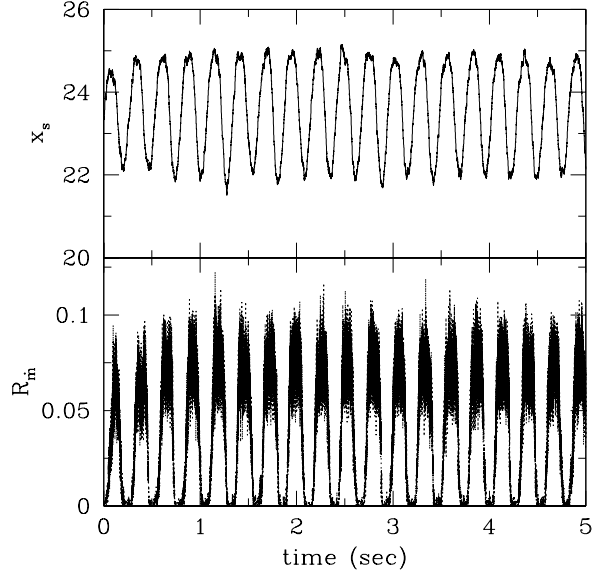}
}
\caption{Variation of shock location ({\it top panel}) and mass outflow
rate ({\it bottom panel}) with time for $10M_\odot$ black hole. Input
parameters are $x_{\rm inj} = 50.4$, $v_{\rm inj} = 0.0643$,
$a_{\rm inj} = 0.0633$, $\lambda_{\rm inj} = 1.65$,
$\alpha = 0.006$ and $\gamma = 4/3$, respectively.
}
\end{figure}
\begin{figure}
\centerline{\includegraphics[width=8cm]{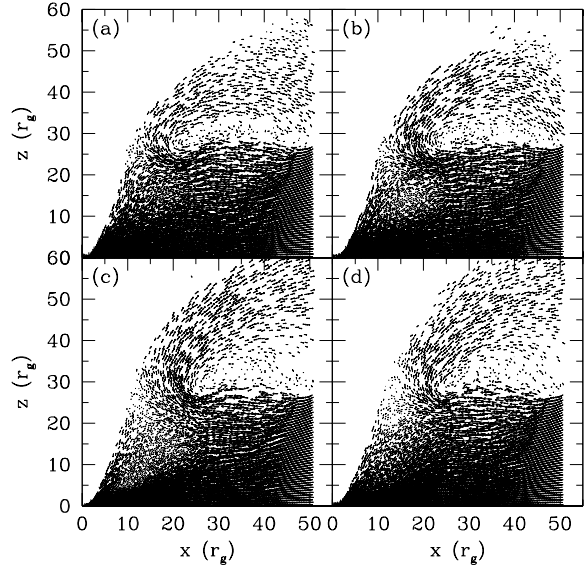}}
\caption{Velocity vectors are in $x-z$ plane. Snap shots are
taken at equal interval within a complete period of shock oscillation.
Input parameters are same as figure 1.}
\end{figure}
\begin{figure}
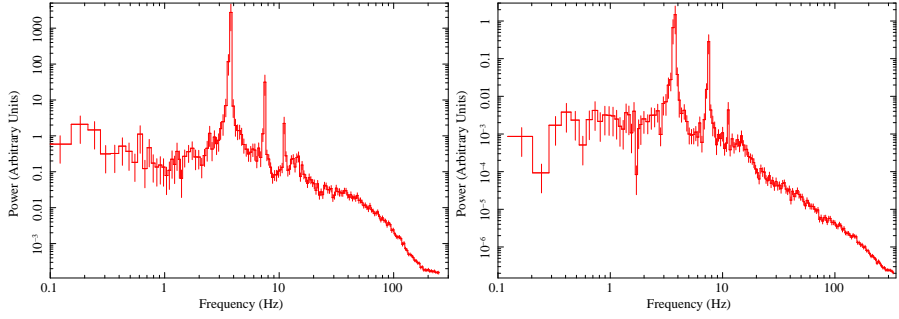

\centerline{\includegraphics[angle=270,width=5.8cm]{Amda1p65_VIS0p006_fort19.ps} 
            \includegraphics[angle=270,width=5.8cm]{Amda1p65-alp0p006-brm-cenbol-fort19.ps}}
\caption{Fourier spectra of shock location variation ({\it left}).
Corresponding power density spectra of bremsstrahlung flux variation ({\it right}).
Input parameters are same as figure 1. See text for details.}
\end{figure}

\section{Results and discussion}

\vskip -0.15in

In our simulation, we inject SPH particles with radial velocity
$v_{\rm inj}$, angular momentum ${\lambda}_{\rm inj}$ and sound speed
$a_{\rm inj}$ from the injection radius $x_{\rm inj}$. At $x_{\rm inj}$,
the disk height is estimated assuming the flow remain in hydrostatic
equilibrium along the vertical direction and obtained as
$H_{\rm inj} \sim a_{\rm inj} x^{1/2}_{\rm inj}(x_{\rm inj}-1)$ \citep{c89}.
During accretion, rotating matter feels centrifugal force that 
opposes the strong pull of gravity close to the black hole. As a result,
matter accumulates there and an effective boundary layer is developed
which is termed as CENtrifugal pressure supported BOundary Layer (CENBOL). 
With the appropriate choice of the outer boundary condition (OBC),
the inviscid accreting flow undergoes shock transition at the CENBOL
and as time evolves, shock becomes stationary. Subsequently, we turn
on the viscosity to transport the
angular momentum outwards which apparently drives the shock front to
recede away from the black hole and finally stabilizes again. For
a given OBC, when the
viscosity is increased further and reached its critical value,
shock front exhibits regular oscillation which sustain forever.
The reason for this resonance oscillation perhaps due to the fact
that the flow jumps from super-sonic branch to 
the sub-sonic branch while attempting to choose the high entropy
solution, but could not satisfy the standing shock conditions and
eventually oscillates.
Due to differential motion, the shock front oscillates like a flap
with respect to disk equatorial plan and the post-shock matter
periodically experiences compression and rarefaction. With the
combined effects of compressional heating and the flapping action
of the shock-front, a part of the post-shock matter is ejected out
from the disk in the vertical direction with modulation at par with
the shock oscillation. When the spewed up matter receives excess driving,
matter leaves the disk in the form of outflow, otherwise falls
back onto the disk. Accordingly, a cycle of periodic mass ejection
is observed from the vicinity of the gravitating object which is
nicely accounted as quasi periodic variation of mass outflow.
In Fig. 1, we present the persistent shock oscillation 
(top panel) taking place over a large period of time for
$10 M_\odot$ black hole. The input parameters are
$x_{\rm inj} = 50.4~{\rm r_g}$, $v_{\rm inj} = 0.0643$,
$a_{\rm inj} = 0.0633$, $\lambda_{\rm inj} = 1.65$,
$\alpha = 0.006$ and $\gamma = 4/3$, respectively.
The corresponding mass outflow rate
[$R_{\dot m}$, defined as the ratio of the outflow rate 
(${\dot M}_{\rm out}$) to the inflow rate 
(${\dot M}_{\rm in}$)]
variation is shown in the lower panel. In Fig. 2, we depict
four snap shots of velocity distribution of SPH particles over
a complete cycle of shock oscillation. It is observed that the
rate of outflow is regulated by the modulation of the post-shock
matter confined within the CENBOL.

\begin{figure}
\centerline{\includegraphics[angle=270,width=8cm]{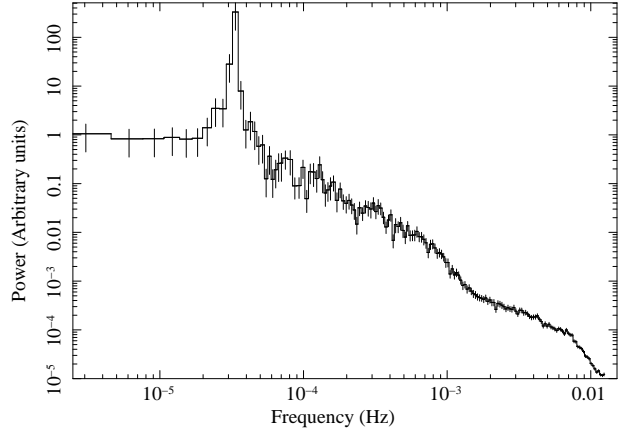} 
}
\caption{Fourier spectra of shock location variation for super massive
black hole of mass $10^6 M_\odot$. Input parameters are $x_{\rm inj} = 50.8$,
$v_{\rm inj} = 0.0653$, $a_{\rm inj} = 0.0622$, $\lambda_{\rm inj} = 1.67$,
$\alpha = 0.00525$ and $\gamma = 4/3$, respectively.}
\end{figure}

Due to shock transition, the post shock matter becomes hot and dense
which would essentially be responsible to emit high energy radiation.
At the critical viscosity, since the shock front exhibits regular
oscillation, the inner part of the disk, $i. e.$ CENBOL, also
oscillates indicating the variation of photon flux emanating from
the disk. Thus, a correlation between the variation of shock front
and emitted radiation seem to be viable. In this work, we estimate
the bremsstrahlung emission as,
$$
E_{Brem} = \int_{x_{1}}^{x_{2}} \rho^2 T^{1/2} x^2 dx,
$$
where, $x_1$ and $x_2$ are the radii of interest within which radiation
is being computed. We calculate the total bremsstrahlung
emission for the matter from the CENBOL 
region. To understand the correlation between the shock oscillation
and the emitted photon flux from the
inner part of the disk, we calculate the Fourier spectra of the
quasi periodic variation of the shock front and the power spectra
of bremsstrahlung flux for matter resides within the boundary of
CENBOL. The obtained results are shown in Fig. 3 where left figure is
for shock oscillation and the right figure is for photon flux
variation. We find that the quasi-periodic variation of the
shock location and the photon flux is characterized by the fundamental
frequency $\nu_{\rm fund} = 3.84$ Hz which is followed by multiple
harmonics. The first few prominent harmonic frequencies are 
7.55 Hz ($\sim 2 \times \nu_{\rm fund}$), 11.45 Hz
($\sim 3 \times \nu_{\rm fund}$) and 14.6 Hz
($\sim 4 \times \nu_{\rm fund}$). This suggests that the dynamics
of the inner part of the disk ($i. e.$ CENBOL) and emitted fluxes
are tightly coupled. The obtained power density spectra (PDS)
of emitted radiation has significant similarity with number of
observational results \citep{cbhs04,ndmc12,nrs13,rn13} and accordingly, this
establishes the fact that the origin of such photon flux
variation seems to be due to the hydrodynamic modulation of the
inner part of the disk in terms of shock oscillation.

In similar context, we study the quasi periodic variation of
shock location around the super massive black hole of mass
$10^6 ~{\rm M_\odot}$. Here, the 
input parameters are chosen as $x_{\rm inj} = 50.8~{\rm r_g}$,
$v_{\rm inj} = 0.0653$, $a_{\rm inj} = 0.0622$, $\lambda_{\rm inj} = 1.67$,
and $\gamma = 4/3$, respectively. We observe that shock starts to 
oscillates for $\alpha = 0.00525$. We calculate the Fourier spectra of
the shock location variation and find that most of the power is
concentrated at a narrow range of frequency $\nu = 3.35 \times 10^{-5} Hz$
which is shown in Fig. 4.
Following the example of stellar mass black hole, we conclude that 
the radiant bremsstrahlung fluxes from the CENBOL matter are also
exhibit quasi periodic oscillation with similar frequency which is
consistent with the recent observational findings 
\citep{cbhs04,ndmc12,nrs13,rn13,in13}. 

\section*{Acknowledgements}

Authors are thankful to Prof. Diego Molteni for sharing the numerical 
code and useful discussion.


\end{document}